\documentstyle[12pt]{article}
\begin{document}
\renewcommand{\thefootnote}{\fnsymbol{footnote}}
\renewcommand{\thesection}{\Roman{section}}
\begin{flushright}
ADP-94-20/T160
\end{flushright}
\vspace{0.5cm}
\begin{center}
\begin{LARGE}
Variations of Hadron Masses and Matter Properties in Dense Nuclear
Matter
\end{LARGE}
\end{center}
\vspace{0.5cm}
\begin{center}
\begin{large}
K.~Saito\footnote{ksaito@nucl.phys.tohoku.ac.jp} \\
Physics Division, Tohoku College of Pharmacy \\ Sendai 981, Japan \\
and  \\
A.~W.~Thomas\footnote{athomas@physics.adelaide.edu.au} \\
Department of Physics and Mathematical Physics \\
University of Adelaide, South Australia, 5005, Australia
\end{large}
\end{center}
\vspace{1.5cm}
\begin{abstract}
Using a self-consistent quark model for nuclear matter we investigate
variations of the masses of the non-strange vector mesons,
the hyperons and the nucleon in
dense nuclear matter (up to four times the normal nuclear density).
We find that the changes in the hadron masses can be described in terms
of the value of the scalar mean-field in matter.  The model is then used to
calculate the density dependence of the quark condensate in-medium,
which turns out to be well approximated by a linear function of the
nuclear density.  Some relations among the hadron properties and the
in-medium quark condensate are discussed.
\end{abstract}
%
%
\newpage
\section{INTRODUCTION}

One of the most interesting future directions in nuclear physics may
be to
study how nuclear matter properties change as the environment
changes. Forthcoming ultra-relativistic heavy-ion experiments are
expected to give significant information on the strong interaction
(i.e., QCD ) in matter, through the detection of changes in hadronic
properties~\cite{hfn,ko,br}.
In particular, the variations in hadron masses in the
nuclear medium have attracted wide interest because such changes
could be a signal of the formation of hot hadronic and/or quark-gluon
matter in the energetic nucleus-nucleus collisions.
Several authors have recently studied the vector-meson ($\omega$,
$\rho$, $\phi$) masses using the vector dominance model~\cite{asak92},
QCD sum rules~\cite{hatsuasa} and the Walecka
model (QHD)~\cite{sa89,will,hatsu94}, and have reported that the mass
decreases in the nuclear medium. There is also a proposal to look for
such mass shifts at CEBAF~\cite{ceb}.

In the approach based on
QCD sum rules, the reduction of
the mass is mainly due to the four-quark condensates and one of the
twist-2 condensates, $\langle \bar{q} \gamma_{\mu} D_{\mu} q
\rangle$.  However, it has been suggested that there may
be considerable, intrinsic
uncertainty in the standard assumptions underlying the QCD sum-rule
analyses~\cite{griegel}.  On the other hand, in hadronic models
like QHD~\cite{walecka}, the main reason for the
reduction in masses is the polarization of the Dirac
sea~\cite{sa89,will,hatsu94},
where the {\em antinucleons} in matter play a crucial role.
However, from the point of view of the quark model, the strong
excitation of {\em nucleon-antinucleon pairs\/} in medium is
difficult to understand.
It is clear that these two mechanisms are quite different.

Several years ago Guichon~\cite{guichon} proposed an entirely
different model for nuclear matter, based on a mean-field description
in which quarks (in nucleon bags) interact self-consistently
with $\sigma$ and
$\omega$ mesons.  The model has been refined by
Fleck {\it et al.}~\cite{fleck} and the present authors.  It provides
a natural explanation of nuclear saturation and the right magnitude
of the nuclear compressibility.  We have used this model to
investigate the nuclear structure functions~\cite{st1}, the
properties of both nuclear and neutron matter~\cite{st2} and the
Okamoto-Nolen-Schiffer anomaly and isospin symmetry breaking in
matter~\cite{st3}.
We argued that the response of the internal structure of the
nucleon to its environment is vital to the understanding of, not only
physics with momentum transfers of several GeV (eg., deep-inelastic
scattering), but also physics at scale of a few MeV (eg., a violation
of charge symmetry in nuclear medium).  In Ref.\cite{st2}, the
relationship between the Guichon model (alias the quark-meson coupling
(QMC) model in our papers) and QHD has also been clarified.    Even
though this model is extremely simple, the insights gained from it may
help to
reveal the essential physics~\cite{ts1}.  Here
we use the Guichon model to investigate variations of various hadronic
properties, as well as the density dependence of the quark condensate
in nuclear matter, as functions of the nuclear density up to
four times
normal nuclear density.

This paper is organized as follows.  In Sec.~II, the QMC model is
introduced and applied to calculate the modification of the
properties of the nucleon in nuclear matter.  The masses of the
vector mesons
($\omega$, $\rho$) and the hyperons ($\Lambda$, $\Sigma$, $\Xi$) in
matter are studied in Sec.~III.
We find that the change in the hadron masses can be described in terms
of scalar mean-field values in medium.  Some relationships among the
hadron masses are given.
Next, in Sec.~IV, we use this model to study the density dependence of
the quark condensate, which may be linked to a diverse range of
nuclear phenomena.
The density dependence of the quark condensate can be well fitted by a
linear function of the nuclear density.  Then, some connections among
the hadron properties and the quark condensate in medium are
discussed.  Sec.~V contains our conclusions.

\section{MATTER PROPERTIES IN DENSE MEDIUM}
\subsection{The quark-meson coupling model}

The QMC model treats nuclear matter as a collection of (nucleon)
MIT bags, self-consistently bound by the exchange of scalar ($\sigma$)
and vector ($\omega$, $\rho$) mesons.  We assume that nuclear matter
with $N \ne Z$ is uniformly distributed, and that the mesons can be
treated
in mean-field approximation (MFA).  Let the mean-field values for
the $\sigma$, $\omega$ (the time
component) and $\rho$ (the time component in the third direction of
isospin) fields be $\bar{\sigma}$, $\bar{\omega}$ and $\bar{b}$,
respectively.  The quarks in a static spherical bag interact with
those mean fields.  The Dirac equation for
a quark field, $\psi_q$, in a bag is then given by
\begin{equation}
[i\gamma\cdot\partial - (m_q - V_{\sigma}) - \gamma^0(V_{\omega} +
\frac{1}{2}\tau_zV_{\rho})]
\psi_q = 0, \label{eq:dirac}
\end{equation}
where $V_{\sigma}=g_{\sigma}^{q}\bar{\sigma}$, $V_{\omega}=
g_{\omega}^{q}\bar{\omega}$ and $V_{\rho}=g_{\rho}^q\bar{b}$, with the
quark-meson coupling
constants, $g_{\sigma}^q$, $g_{\omega}^q$ and $g_{\rho}^q$.  The bare
quark mass is
denoted by $m_q$ and
$\tau_z$ is the third Pauli matrix.  The normalized,
ground state
for a quark is then given by
\begin{equation}
\psi_q(\vec{r},t) = {\cal N}_q \exp[-i\epsilon_q t/R]
{j_{0}(x_q r/R) \choose
i\beta_q {\vec{\sigma}}\cdot\hat{r}j_{1}(x_q r/R)}
{\frac{\chi_q}{\sqrt{4\pi}}}, \label{eq:psiq}
\end{equation}
where
\begin{equation}
\epsilon_q = \Omega_q + R (V_{\omega} \pm \frac{1}{2}
V_{\rho}), \mbox{ for } {u \choose d} \mbox{ quark,} \label{eq:epq}
\end{equation}
\begin{equation}
{\cal N}_q^{-2} = 2R^3j^2_0(x_q)[\Omega_q(\Omega_q - 1)
+
R m_q^{\star}/2]/x_q^2, \label{eq:norm}
\end{equation}
\begin{equation}
\beta_q = \sqrt{(\Omega_q - R m_q^{\star})/(\Omega_q +
R m_q^{\star})}, \label{eq:betq}
\end{equation}
with $\Omega_q = \sqrt{x_q^2 + (R m_q^{\star})^2}$ and $\chi_q$
the
quark spinor.  The effective quark mass, $m_q^{\star}$, is defined
by
\begin{equation}
m_q^{\star} = m_q - V_{\sigma}, \label{eq:qem}
\end{equation}
for both {\it u\/} and {\it d\/} quarks.
The linear boundary condition, $j_0(x_q) = \beta_q j_1(x_q)$, at the
bag
surface determines the
eigenvalue, $x_q$.

Using the SU(6) spin-flavor wave function for the nucleon,
the nucleon energy is given by $E_{bag}^N + 3V_{\omega} \pm \frac{1}
{2}V_{\rho}$ for ${p \choose n }$, where the bag energy is
\begin{equation}
E_{bag}^N = {\frac{\sum_q n_q \Omega_q - z_N}{R}} +
{\frac{4}{3}}\pi
BR^3,
\label{eq:bageb}
\end{equation}
with $B$ the bag constant and $z_N$ a phenomenological parameter
accounting for a multitude of corrections, including zero-point motion.
 Here $n_q$ is the number of quarks in the nucleon.
To correct for spurious c.m. motion
in the bag~\cite{boost} the mass of the nucleon at rest is taken to be
\begin{equation}
M_N = \sqrt{(E_{bag}^N)^2 - \sum_q n_q (x_q/R)^2}.
\label{eq:cmc}
\end{equation}
The effective nucleon mass, $M_N^{\star}$, in nuclear matter is given
by minimizing eq.(\ref{eq:cmc}) with respect to $R$.

\begin{table}[hbtp]
\begin{center}
\caption{$B^{1/4}$ and $z_N$ for some bag radii ($m_0=$ 5 MeV).}
\label{bagcon}
\begin{tabular}[t]{cccc}
\hline\hline
 $R_0 (fm)$      & 0.6 & 0.8 & 1.0 \\
\hline
 $B^{1/4}$ (MeV) & 187.7 & 157.2 & 136.1 \\
 $z_N$           & 2.038 & 1.640 & 1.169 \\
\hline\hline
\end{tabular}
\end{center}
\end{table}
To see the sensitivity of our results to the bag radius of the free
nucleon, $R_0$, we choose the current quark mass, $m_q = m_0 (= m_u =
m_d) = 5$ MeV, and vary the parameters, $B$ and $z_N$, to fit the
nucleon mass at 939 MeV with $R_0$ (= 0.6, 0.8, 1.0 $fm$).
The values of $B^{1/4}$ and $z_N$ are listed in Table~\ref{bagcon}.

For infinite nuclear matter we take the Fermi momenta for protons and
neutrons to be $k_{F_j}$ ($j=p$ or $n$). This is
defined by $\rho_j = k_{F_j}^3 / (3\pi^2)$, where $\rho_j$ is the
density of
protons or neutrons, and the total baryon density,
$\rho_B$, is then given by $\rho_p + \rho_n$.

Since we want to calculate the variations in the masses of the $\omega$
and $\rho$ mesons, we suppose that both mesons are also described by
the MIT bag model in the scalar mean-field.  To fit the free masses,
$m_{\omega}$ = 783 MeV and $m_{\rho}$ = 770 MeV, we introduce new
$z$-parameters for the two mesons, $z_{\omega}$ and $z_{\rho}$.  We
then find that $z_{\omega(\rho)}$ = 0.7840(0.8061), 0.4812(0.5160),
0.1198(0.1680) for $R_0$ = 0.6, 0.8, 1.0 $fm$, respectively.
Unfortunately, in this model it is hard to deal with the density
dependence of the $\sigma$ meson in medium because it couples strongly
to the pseudoscalar ($\pi$) channel, which requires a direct
treatment of chiral symmetry in medium~\cite{hatkun}.  That is
beyond the scope of this study.  Though one might expect the
$\sigma$-meson mass in medium to be less than the free one (see, eg.,
Ref.~\cite{sa89}), we shall keep the free value,
550 MeV, even in medium.

The $\omega$ field is now
determined by baryon number conservation as $\bar{\omega}=g_{\omega}
\rho_B / m_{\omega}^{\star 2}$, where $g_{\omega} = 3g_{\omega}^q$ and
$m_{\omega}^{\star}$ is the effective $\omega$-meson mass,
and the $\rho$ mean-field by the difference in proton and neutron
densities, $\rho_3 = \rho_p - \rho_n$. On the other hand, the scalar
mean-field is given by a self-consistency
condition (SCC)~\cite{walecka,st2}.  Since the $\rho$ field value is
$\bar{b} = g_{\rho} \rho_3 / (2m_{\rho}^{\star 2})$, where
$g_{\rho}=g_{\rho}^q$, the total
energy per nucleon, $E_{tot}$, can be written
\begin{equation}
E_{tot} = \frac{2}{\rho_B (2\pi)^3}\sum_{j=p,n}\int^{k_{F_j}}
d\vec{k} \sqrt{M_N^{\star 2} + \vec{k}^2} + \frac{m_{\sigma}^2}
{2\rho_B}{\bar{\sigma}}^2 + \frac{g_{\omega}^2}
{2m_{\omega}^{\star 2}}\rho_B + \frac{g_{\rho}^2}{8m_{\rho}^{\star 2}
\rho_B}
 \rho_3^2, \label{eq:tote}
\end{equation}
where $m_{\rho}^{\star}$ is the effective $\rho$-meson mass in medium.
Then, the SCC for the $\sigma$ field is given by
\begin{equation}
\bar{\sigma} = - \frac{2}{(2\pi)^3 m_{\sigma}^2} \left[ \sum_{j=p,n}
\int^{k_{F_j}} d\vec{k} \frac{M_j^{\star}}
{\sqrt{M_j^{\star 2} + \vec{k}^2}} \right]  \left(\frac{\partial
M_N^{\star}}{\partial \bar{\sigma}}\right)_{R}.
\label{eq:scc}
\end{equation}
Using Eqs.(\ref{eq:bageb}) and (\ref{eq:cmc}), we find
\begin{equation}
\left(\frac{\partial M_N^{\star}}{\partial \bar{\sigma}}
\right)_{R} \equiv - g_{\sigma}
C_N(\bar{\sigma}), \label{eq:ders}
\end{equation}
where $g_{\sigma}=3g_{\sigma}^q$ and
$C_N$ is the quark-scalar density in the nucleon:
\begin{equation}
C_N(\bar{\sigma}) = \left( \frac{E_{bag}^N}{M_N^{\star}} \right)
\left[ \left( 1-\frac{\Omega_q}{E_{bag}^N R} \right) S_N +
\frac{m_q^{\star}}
{E_{bag}^N} \right], \label{eq:sd}
\end{equation}
with
\begin{equation}
S_N = \int_R d{\vec r} {\bar{\psi}}_q \psi_q =
\frac{\Omega_q /2+R m_q^{\star}(\Omega_q -1)}{\Omega_q
(\Omega_q -1)+R m_q^{\star}/2}. \label{eq:qsd}
\end{equation}
The value of the quark-scalar density is about $0.4$ at $\rho_B = 0$,
and it decreases monotonically to about $0.2$ at the normal nuclear
density, $\rho_0 = 0.17 fm^{-3}$.  The dependence of the scalar
density on the bag radius is not strong (see Ref.\cite{st2}).

\subsection{Coupling constants and nuclear matter properties}

We determine the coupling constants, $g_{\sigma}^2$ and
$g_{\omega}^2$, so as to fit the binding energy ($-16$ MeV) at the
saturation density, $\rho_0$, for symmetric
nuclear matter. Furthermore, the $\rho$-meson coupling constant is
used to reproduce the bulk symmetry energy, $33.2$ MeV.
\begin{table}[hbtp]
\begin{center}
\caption{The coupling constants and calculated properties of
equilibrium matter at the saturation density ($m_0$ = 5 MeV).  The
effective nucleon mass, $M_N^{\star}$, and the nuclear
compressibility, $K$, are quoted in MeV.}
\label{coupc}
\begin{tabular}[t]{cccccccc}
\hline\hline
 $R_0 (fm)$ & $g_{\sigma}^2/4\pi$ & $g_{\omega}^2/4\pi$ &
$g_{\rho}^2/4\pi$ & $M_N^{\star}$ & $K$ & $\frac{\delta R}{R_0}$ &
$\frac{\delta x}{x_0}$ \\
\hline
 0.6        & 18.79 & 1.326 & 4.923 & 838 & 223 & -0.03 & -0.07 \\
 0.8        & 20.63 & 1.016 & 5.014 & 850 & 205 & -0.02 & -0.09 \\
 1.0        & 21.09 & 0.871 & 5.045 & 855 & 196 & -0.01 & -0.12 \\
\hline\hline
\end{tabular}
\end{center}
\end{table}
The values of the coupling constants are listed in Table~\ref{coupc}.
In the last two columns, the relative changes (with respect to the
values at zero density) of the bag radius and the lowest eigenvalue
are shown.  The present model gives a good value for the nuclear
compressibility -- around 200 MeV.  If we take a heavier current-quark
mass, the effective nucleon mass and the nuclear compressibility are,
respectively, reduced and enhanced.
\begin{figure}[hbt]
\begin{center}
\caption{Scalar mean-field values.
The dotted, solid and dashed curves are, respectively, for $R_0$ =
0.6, 0.8 and 1.0 $fm$.}
\label{vsig}
\end{center}
\end{figure}
The strength of the scalar mean-field, $V_{\sigma}$, in medium is
shown in Fig.\ref{vsig}.  At small density it is well approximated by
a linear function of the density (see also Eq.(40) in Ref.\cite{st2}):
\begin{equation}
V_{\sigma} \approx 140 \mbox{ (MeV) } \left( \frac{\rho_B}{\rho_0}
\right). \label{eq:appv}
\end{equation}
\begin{figure}[ht]
\begin{center}
\caption{Effective nucleon mass in symmetric nuclear matter.
The curves are labelled as in Fig.1.}
\label{efnms}
\end{center}
\end{figure}
The dependence of the effective nucleon mass on the nuclear density
is shown in Fig.\ref{efnms}.  It decreases as the density goes up, and
it behaves like constant at large density.  We find that the effective
nucleon mass at small density is approximately given by
\begin{equation}
\frac{M_N^{\star}}{M_N} \simeq 1 - 0.14 \left( \frac{\rho_B}{\rho_0}
\right), \label{eq:nms}
\end{equation}
which is identical to the model independent result derived using QCD
sum-rules~\cite{druka}.

\begin{figure}[hbt]
\begin{center}
\caption{The ratio of the axial-vector coupling constant of the
nucleon in medium to that in free space.  The curves are labelled as
in Fig.1.}
\label{gastr}
\end{center}
\end{figure}
\begin{figure}[ht]
\begin{center}
\caption{The ratio of the magnetic moment of the proton in medium to
that in free space.  The curves are labelled as in Fig.1.}
\label{magmom}
\end{center}
\end{figure}
In Fig.\ref{gastr} the density dependence of the axial-vector
coupling constant of the nucleon is illustrated.  It decreases as the
density increases.  On the other hand, the magnetic moment of the proton
in symmetric nuclear matter increases with density as shown
in Fig.\ref{magmom}. The calculation
of these quantities is standard~\cite{mit} (c.m. and recoil corrections
are not included~\cite{g83}).  It is rather easy to understand why
$g_A^{\star}$ and $\mu_N^{\star}$ tend to decrease and increase,
respectively, with density: in the bag model these quantities involve
integrals over the upper and lower components of the quark wave
function.  Because the attractive scalar potential effectively
decreases the quark mass it makes the solution more relativistic --
i.e., the lower component of the wave function is enhanced.  This
fact leads, respectively, to the decrease and increase of $g_A^{\star}$
and $\mu_N^{\star}$ in medium.  Given the simplicity of this argument
we find the conclusion that the magnetic moments of the other hadrons
should increase quite compelling.

At small density, we can expand $g_A^{\star}$ and $\mu_N^{\star}$ in
terms of $\rho_B$. If we take the quark to be massless for simplicity,
and ignore the dependence of $g_A^{\star}$ and
$\mu_N^{\star}$ on the change of the bag radius in medium, they are
expressed as
\begin{equation}
\frac{g_A^{\star}}{g_A} \simeq 1 - \frac{4x_0^3-12x_0^2+10x_0-3}
{2x_0^2(x_0-1)^2} (RV_{\sigma}) \simeq 1 - 0.09 \left( \frac{\rho_B}
{\rho_0} \right), \label{eq:axst}
\end{equation}
and
\begin{equation}
\frac{\mu_N^{\star}}{\mu_A} \simeq 1 - \frac{4x_0^3-16x_0^2+17x_0-6}
{2x_0(x_0-1)^2(4x_0-3)} (RV_{\sigma}) \simeq 1 + 0.1 \left(
\frac{\rho_B}{\rho_0} \right), \label{eq:magst}
\end{equation}
where $x_0$ = 2.04, $R$ = 0.8 $fm$ and Eq.(\ref{eq:appv}) is used.
These formulae can roughly reproduce both the ratios at small density.
Furthermore, we have calculated the charge radius of the proton in
medium, and we find that the change is within a few per cent even at
$\rho_B \sim 4 \rho_0$.

Here we must add a caution that the contribution of meson exchange
currents (MEC) to $g_A^{\star}$, $\mu_N^{\star}$, etc. would be
considerable in medium, and, hence, we should treat the whole problem
including MEC to get the final results for the changes in these
quantites. This is obviously beyond the scope of the present work.

Finally, we record that the variations of the above nucleon
properties depend only very weakly on the proton fraction, $f_p$,
which is
defined as $\rho_p / \rho_B$, because the $\rho$ meson does not play
a role in determining the nucleon structure in medium (see
Eqs.(\ref{eq:qem})~$\sim$~(\ref{eq:cmc})).  Further investigations
concerning the nucleon properties, the equation of state for matter,
etc. can be found in Ref.\cite{st1,st2,st3}.
\clearpage

\section{HADRON MASSES IN MEDIUM}

Now we are in a position to present our main results.
\begin{figure}[ht]
\begin{center}
\caption{The ratio of the vector-meson mass in symmetric nuclear
matter to the free mass.  The curves are labelled as in Fig.1.}
\label{vecms}
\end{center}
\end{figure}
In Fig.\ref{vecms}, the ratio of the effective vector-meson mass,
$m_v^{\star}$, to that in free space is shown as a function of the
density.  Since the difference between the effective $\omega$- and
$\rho$-meson masses at the same density is very small, we show only
one curve for both mesons in the figure.  As the density increases
the vector-meson mass decreases (as several authors have previously
noticed~\cite{asak92,hatsuasa,sa89,hatsu94}), and seems to become
flat like the effective nucleon mass.  The mass reduction can be
well expressed by a linear form at small density:
\begin{equation}
\frac{m_v^{\star}}{m_v} \simeq 1 - 0.09 \left( \frac{\rho_B}{\rho_0}
\right). \label{eq:vmms}
\end{equation}
The reduction factor, 0.09, is somewhat smaller than that predicted
by the QCD sum rules~\cite{hatsu94}.  In this model the reduction in
the mass is basically caused by the attractive scalar mean-field in
medium, and this origin is clearly different from that in
QHD~\cite{sa89,hatsu94}, in which the essential mechanism is
the vacuum
polarization due to the excitation of {\it nucleon-antinucleon pairs\/}
 in medium.

It is possible to calculate masses of other hadrons in this model.
In particular, there is considerable interest in studying the properties
of hyperons in
medium -- eg. $\Lambda$, $\Sigma$ and $\Xi$.  For the hyperons
themselves we again use the MIT bag model, including the c.m.
corrections (see Eq.(\ref{eq:cmc})).  We assume that the strange
quark does not couple to the $\sigma$ meson in MFA, and that the
addition of a single hyperon to nuclear matter of density $\rho_B$
does not alter the values of the scalar and vector mean-fields.  The
mass
of the strange quark, $m_s$, is determined so as to reproduce the mass
splitting between the nucleon and the average of $\Lambda$ and
$\Sigma$ hyperons, $M_{H_{av}} (= 1154$ MeV), in free space.  We then
find $m_s$ = 355.5, 358.1, 354.0 MeV for $R_0$ = 0.6, 0.8, 1.0
$fm$, respectively.
Using these values of the strange-quark mass, we have calculated the
average mass of $\Lambda$ and $\Sigma$, $M_{H_{av}}^{\star}$, and the
mass of $\Xi$, $M_{\Xi}^{\star}$, in symmetric nuclear matter.

\begin{figure}[hbt]
\begin{center}
\caption{The ratio of the average mass of $\Lambda$ and $\Sigma$ in
medium to that in free space.  The curves are labelled as in Fig.1.}
\label{lamsig}
\end{center}
\end{figure}
In Fig.\ref{lamsig}, the average mass, $M_{H_{av}}^{\star}$, in medium
is illustrated.  The density dependence of $M_{H_{av}}^{\star}$ is
very similar to the cases of the vector meson and the nucleon.  The
ratio is again well described by a linear function at small density:
\begin{equation}
\frac{M_{H_{av}}^{\star}}{M_{H_{av}}} \simeq 1 - 0.08 \left(
\frac{\rho_B}{\rho_0} \right). \label{eq:avms}
\end{equation}
Note that the reduction factor in the mass formula is almost the same
value as in Eq.(\ref{eq:vmms}).

\begin{figure}[ht]
\begin{center}
\caption{The ratio of the $\Xi$ mass in medium to that in free space.
The curves are labelled as in Fig.1.}
\label{xi}
\end{center}
\end{figure}
The effective $\Xi$-hyperon mass in medium is shown in Fig.\ref{xi}.
The ratio again behaves like the case of $M_{H_{av}}^{\star}$, but an
approximate form for $M_{\Xi}^{\star} / M_{\Xi}$ at low density is
given by
\begin{equation}
\frac{M_{\Xi}^{\star}}{M_{\Xi}} \simeq 1 - 0.04 \left(
\frac{\rho_B}{\rho_0} \right), \label{eq:xims}
\end{equation}
where the reduction factor is about half of those in
Eqs.(\ref{eq:vmms}) and (\ref{eq:avms}).

\begin{figure}[hbt]
\begin{center}
\caption{The ratios of the effective masses of the hadrons in
symmetric nuclear matter to those in free space ($R_0$ = 0.8 $fm$).
The dashed, solid, dotted and dot-dashed curves are for $\Xi$,
($\Lambda+\Sigma$)/2, the vector ($\omega$, $\rho$) meson and the
nucleon, respectively.}
\label{all}
\end{center}
\end{figure}
We summarize the effective masses of the hadrons in medium in
Fig.\ref{all}.  As one can see from the figure,
the reduction in both the vector-meson mass and the average mass of
$\Lambda$ and $\Sigma$ are about twice that in the $\Xi$-hyperon mass
over a wide range of $\rho_B$, while
the reduction in the nucleon is about three times that in the $\Xi$.
One can also see this fact from the reduction factors in
Eq.(\ref{eq:nms}) and Eqs.(\ref{eq:vmms})~$\sim$~(\ref{eq:xims}).  It
is rather easy to understand why such a relationship holds among the hadron
masses: the nucleon consists of three non-strange quarks, which
couple to the scalar mean-field in medium, while the vector mesons,
and the
$\Lambda$ and $\Sigma$ hyperons, involve just two non-strange quarks.
The $\Xi$ hyperon consists of only one non-strange quark and two
strange quarks.  Since the strange quark does not feel the scalar
mean-field, its mass is not altered in medium.
The change in the hadron mass is therefore roughly determined by the
number of non-strange quarks, $n_0$, and the strength of the scalar
mean-field, $V_{\sigma}$.  We can then find an approximate form of the
hadron mass in symmetric nuclear matter:
\begin{equation}
\frac{M_{hadron}^{\star}}{M_{hadron}} \simeq 1 - \mbox{ const } \times
n_0 V_{\sigma} \mbox{ (MeV)}, \label{eq:appms}
\end{equation}
where const $\simeq 3.2 \times 10^{-4} \mbox{MeV}^{-1}$.  Note that
this formula is not too sensitive to the bag radius, and that it can
reproduce the hadron masses reasonably well over the range of $\rho_B$
up to about $3 \rho_0$. From
Eq.(\ref{eq:appms}) we expect the following relations among the
hadron masses:
\begin{equation}
\left( \frac{M_{N}^{\star}}{M_{N}} \right) \approx \left(
\frac{M_{\Xi}^{\star}}{M_{\Xi}} \right)^3, \hspace{0.5cm} \left(
\frac{m_{v}^{\star}}{m_{v}} \right) \approx \left(
\frac{M_{\Lambda}^{\star}}{M_{\Lambda}} \right) \approx \left(
\frac{M_{\Sigma}^{\star}}{M_{\Sigma}} \right) \approx \left(
\frac{M_{\Xi}^{\star}}{M_{\Xi}} \right)^2. \label{eq:rela}
\end{equation}
Furthermore, using Eqs.(\ref{eq:axst}) and (\ref{eq:magst}),
$g_A^{\star}$, $\mu_N^{\star}$ and $m_v^{\star}$ at small density
could be linked as
\begin{equation}
\left( \frac{m_{v}^{\star}}{m_{v}} \right) \approx \left(
\frac{g_A^{\star}}{g_A} \right) \approx \left( \frac{\mu_{N}^{\star}}
{\mu_{N}} \right)^{-1}. \label{eq:rela2}
\end{equation}
Finally, we record that the proton-fraction dependence of the
effective masses we have studied in this section is again very weak.
\clearpage

\section{QUARK CONDENSATES IN MEDIUM}

Having shown that the QMC model provides an interesting description of
the hadron mass in nuclear matter we now apply it to study the
behavior of the quark condensates in medium.  The quark condensates
are very important parameters in the QCD sum rules, and it is believed
that they are linked to a wide range of nuclear phenomena, including
the effective hadron mass in medium~\cite{br,qcdsum}.

The difference of the quark condensate
in matter, $Q(\rho_B)$, and that in vacuum, $Q(0)$, is given through
the Hellmann-Feynman
theorem~\cite{st2,st3,ts1,cohen}:
\begin{eqnarray}
Q(\rho_B) - Q(0) &=& \frac{1}{2} \frac{\partial {\cal E}}{\partial
m_q}, \nonumber \\
&=& \frac{\partial {\cal E}}{\partial M_N^{\star}} \frac{d M_N^{\star}}
{d m_q} + \sum_{j= mesons} \frac{\partial {\cal E}}{\partial
m_j^{\star}} \frac{d m_j^{\star}}{d m_q} +  \sum_j \frac{\partial
{\cal E}}{\partial g_j} \frac{d g_j}{d m_q} + ... , \label{eq:cond}
\end{eqnarray}
where ${\cal E} = \rho_B E_{tot}$.

Using the SCC, one finds
\begin{equation}
\frac{\partial {\cal E}}{\partial M_N^{\star}} = \frac{1}
{C_N(\bar{\sigma})} \left( \frac{m_{\sigma}}{g_{\sigma}} \right)^2
(g_{\sigma} \bar{\sigma}), \label{eq:dern}
\end{equation}
\begin{equation}
\frac{d M_N^{\star}}{d m_q} = 3C_N(\bar{\sigma}) \left( 1 -  \frac{d
V_{\sigma}}{d m_q} \right), \label{eq:dernq}
\end{equation}
and
\begin{equation}
\frac{d m_v^{\star}}{d m_q} = 2C_v(\bar{\sigma}) \left( 1 -  \frac{d
V_{\sigma}}{d m_q} \right), \label{eq:dervq}
\end{equation}
where $v = \omega$ or $\rho$, and $C_v$ is the quark-scalar density in
the vector meson given by Eq.(\ref{eq:sd}) (using the variables for
the vector meson instead of those for the nculeon).  Then, using the
parametrization for the derivative of the $\sigma$-meson mass with
respect to the quark mass~\cite{cohen}, $\frac{d m_{\sigma}}{d m_q} =
\frac{\sigma_N m_{\sigma}}{m_q M_N}$, where $\sigma_N (=3m_q C_N(0))$
is the nucleon $\sigma$ term~\cite{sigma}, and the
Gell-Mann--Oakes--Renner relation, the ratio of the quark condensate
in medium to that in vacuum is written by a rather lengthy formula:
\begin{equation}
\frac{Q(\rho_B)}{Q(0)} = 1 - \frac{\sigma_N}{C_N(0) m_{\pi}^2
f_{\pi}^2} \left[ \left(\frac{m_{\sigma}}{g_{\sigma}} \right)^2
(g_{\sigma} \bar{\sigma}) + A_1 {\bar{\sigma}}^2 + A_2 \rho_B^2 + A_3
\rho_3^2 \right], \label{eq:cond1}
\end{equation}
where $m_{\pi}$ is the pion mass (138 MeV), $f_{\pi}$ the pion decay
constant (93 MeV) and the coefficients, $A_1 \sim A_3$, are given by
\begin{equation}
A_1 = C_N(0) \frac{m_{\sigma}^2}{M_N} - \frac{1}{6} \left(
\frac{m_{\sigma}}{g_{\sigma}} \right)^2 \frac{d g_{\sigma}^2}{d m_q},
\label{eq:a1}
\end{equation}
\begin{equation}
A_2 = \frac{1}{6 m_{\omega}^{\star 2}} \frac{d g_{\omega}^2}{d m_q}
- \frac{2 C_{\omega}(\bar{\sigma}) g_{\omega}^2}{3
m_{\omega}^{\star 3}} \left( 1- \frac{d V_{\sigma}}{d m_q} \right)
, \label{eq:a2}
\end{equation}
and
\begin{equation}
A_3 = \frac{1}{24 m_{\rho}^{\star 2}} \frac{d g_{\rho}^2}{d m_q}
- \frac{2 C_{\rho}(\bar{\sigma}) g_{\rho}^2}{12 m_{\rho}^{\star 3}}
\left( 1- \frac{d V_{\sigma}}{d m_q} \right). \label{eq:a3}
\end{equation}
We use $\sigma_N$ = 45 MeV~\cite{sigma}.  The derivatives of the
scalar mean-field and the coupling constants with respect to the
quark mass are calculated numerically  -- eg., the coupling constants
are approximately given by quadratic or linear functions of the quark
mass: for $R_0$ = 0.8 $fm$, $g_{\sigma}^2 = 269.6 - 2.150 m_0 +
0.01438 m_0^2$, $g_{\omega}^2 = 11.97 + 0.1616 m_0$, $g_{\rho}^2 =
63.38 - 0.076 m_0$.

\begin{figure}[ht]
\begin{center}
\caption{The ratio of the quark condensate in medium to that in
vacuum ($m_0$ = 5 MeV and $R_0$ = 0.8 $fm$).
 The solid and dotted curves are, respectively, for $f_p$ = 0.5 and
0.}
\label{cond}
\end{center}
\end{figure}
In Fig.\ref{cond} we show the ratio of the quark condensate in medium
to that
in vacuum.  As one can see, the density dependence of the
quark condensate can be well fitted by a linear function of the
density :
\begin{equation}
\frac{Q(\rho_B)}{Q(0)} \simeq 1 - {0.36 \choose 0.34} \left(
\frac{\rho_B}{\rho_0} \right), \label{eq:condapp}
\end{equation}
where the reduction factors, ${0.36 \choose 0.34}$, are for the
proton fractions, $f_p = {0.5 \choose 0}$, respectively.  The value
of 0.36 agrees with the model-independent prediction of
Cohen {\it et al.}~\cite{cohen}
for symmetric nuclear matter.
Note that the ratio is not sensitive to the
bag radius.  If we take a heavier current quark mass (eg., $m_0$ = 10
MeV), the reduction in the ratio becomes smaller (about 1 \% at
$\rho_B = 2 \rho_0$) than the present case.

Taking up the terms of ${\cal O}(\rho_B)$ in Eq.(\ref{eq:cond}) to
see the behavior at low density, the ratio is given by
\begin{equation}
\frac{Q(\rho_B)}{Q(0)} \simeq 1 - \frac{3 \sigma_N}{C_N(0) m_{\pi}^2
f_{\pi}^2} \left(\frac{m_{\sigma}}{g_{\sigma}} \right)^2 V_{\sigma}
(\bar{\sigma}). \label{eq:cond2}
\end{equation}
Using Eq.(\ref{eq:appv}), $g_{\sigma}^2 \simeq$ 260 and $C_N(0)
\simeq$ 0.37, Eq.(\ref{eq:cond2}) can reproduce Eq.(\ref{eq:condapp}).
 On the other hand, since the ratio of the hadron mass in medium to
that in free space is roughly described by Eq.(\ref{eq:appms}), the
ratio of the quark condensates could be linked to the ratio of the
hadron masses at low density:
\begin{equation}
\frac{M_{hadron}^{\star}}{M_{hadron}} \simeq 1 - 0.124 \times n_0
\left( 1 -
\frac{Q(\rho_B)}{Q(0)} \right). \label{eq:mscond}
\end{equation}
This relation suggests that the ratio of the effective hadron mass to
the free one does not always scale as $\left( \frac{Q(\rho_B)}{Q(0)}
\right)^{1/3}$~\cite{br}.  At small density we could expect
Eq.(\ref{eq:rela}), (\ref{eq:rela2}) and
\begin{equation}
\left( \frac{m_{v}^{\star}}{m_{v}} \right) \approx \left(
\frac{Q(\rho_B)}{Q(0)} \right)^{1/4}, \label{eq:rela3}
\end{equation}
rather than the naive $\left( \frac{Q(\rho_B)}{Q(0)} \right)^{1/3}$
scaling.

\section{CONCLUSION}

We have applied the quark-meson coupling (QMC) model to investigate
the density dependence of the properties of hadrons and the quark
condensates in dense nuclear matter (up to four times the normal
nuclear density).  We have calculated not only the variations in the
nucleon properties in medium but also those in the masses of the
vector ($\omega$, $\rho$) mesons and the hyperons ($\Lambda$,
$\Sigma$, $\Xi$).  As several authors have
suggested~\cite{asak92,hatsuasa,sa89,hatsu94}, the hadron mass is
reduced due to the change of the scalar mean-field in medium.  In the
present model the hadron mass can be related to the number of
non-strange quarks and the strength of the scalar mean-field.  The
hadron masses are simply connected to one another, and the
relationship among them is given by  Eq.(\ref{eq:rela}),
which is effective over a wide range of the nuclear density.
Furthermore, the ratio of the quark condensate in medium to that in
vacuum can be
related to the vector-meson mass and the nucleon
properties -- eg., $g_A^{\star}$ and $\mu_N^{\star}$ -- in medium.

Finally, we would like to give a few caveats concerning the present
calculation.  The basic idea of the model is that the mesons are
locally coupled to the quarks.  This is certainly common for pions in
models like the chiral or cloudy bag~\cite{cloudy}, but may be less
justified for heavier mesons like the vector mesons, which are
obviously
not collective states.  The model also omits the effect of
short-range correlations among the quarks.  At very high density these
would be expected to dominate.  Furthermore, the pionic cloud of the
hadron~\cite{cloudy} should be considered explicitly in any truly
quantitative study of the hadron properties in medium.  As we
mentioned at the end of Sec.~II, the contribution of meson exchange
currents in medium is also important and should be considered in a
consistent manner.  Finally, we note that subtleties such as
scalar-vector mixing in medium and the splitting between longitudinal
and transverse masses of the vector mesons~\cite{will} have been ignored
in the present mean-field study. Although the former appears to be quite
small in QHD the latter will certainly be important in any attempt to
actually measure the mass shift.

\vspace{0.5cm}
We would like to thank A.~G.~Williams for helpful comments on the
manuscript.
This work was supported by the Australian Research Council.
%
%

%
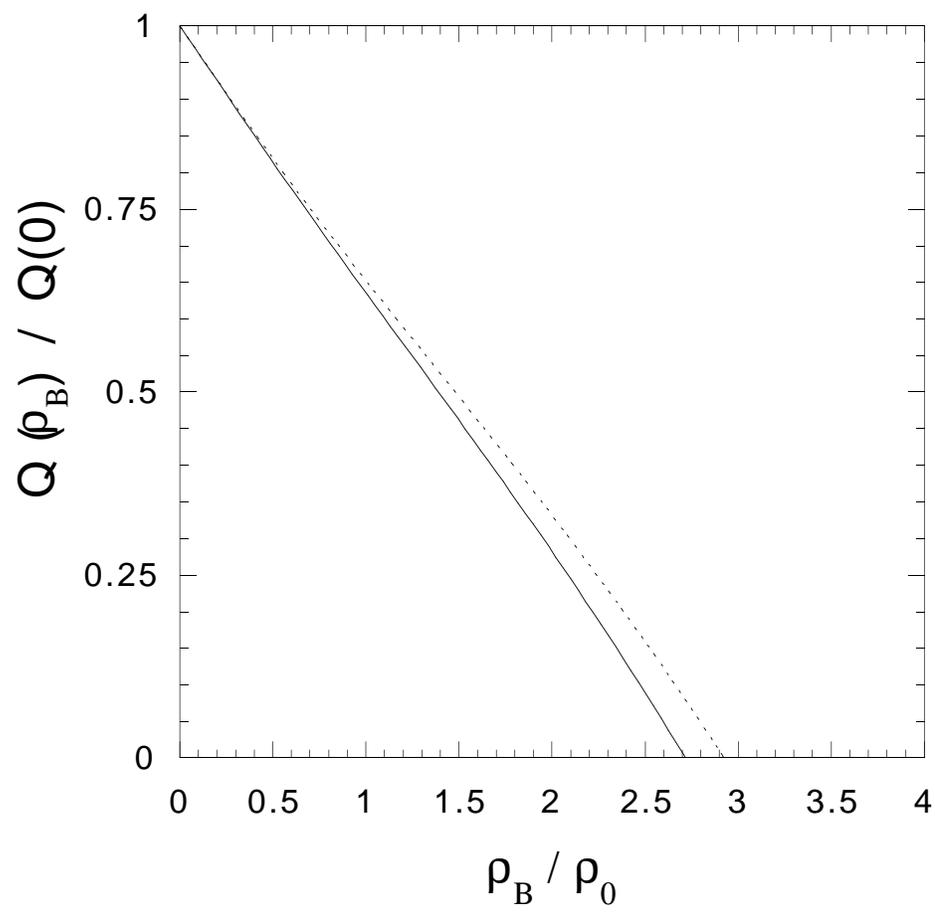
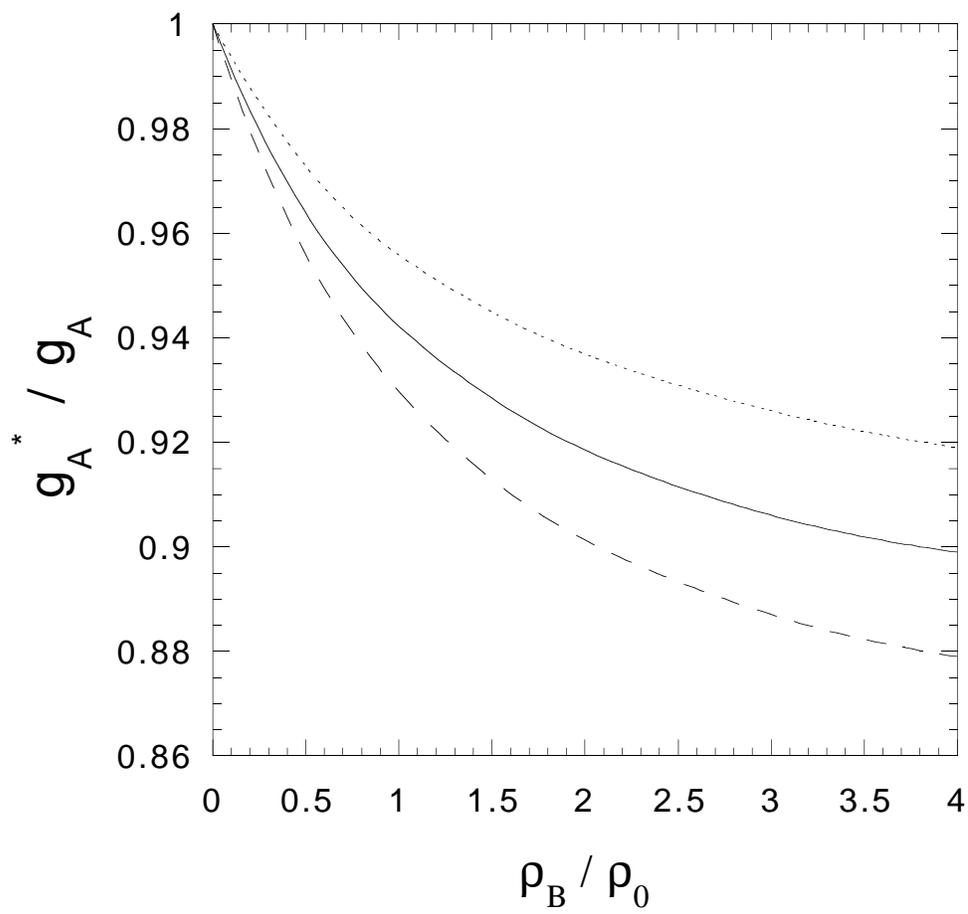
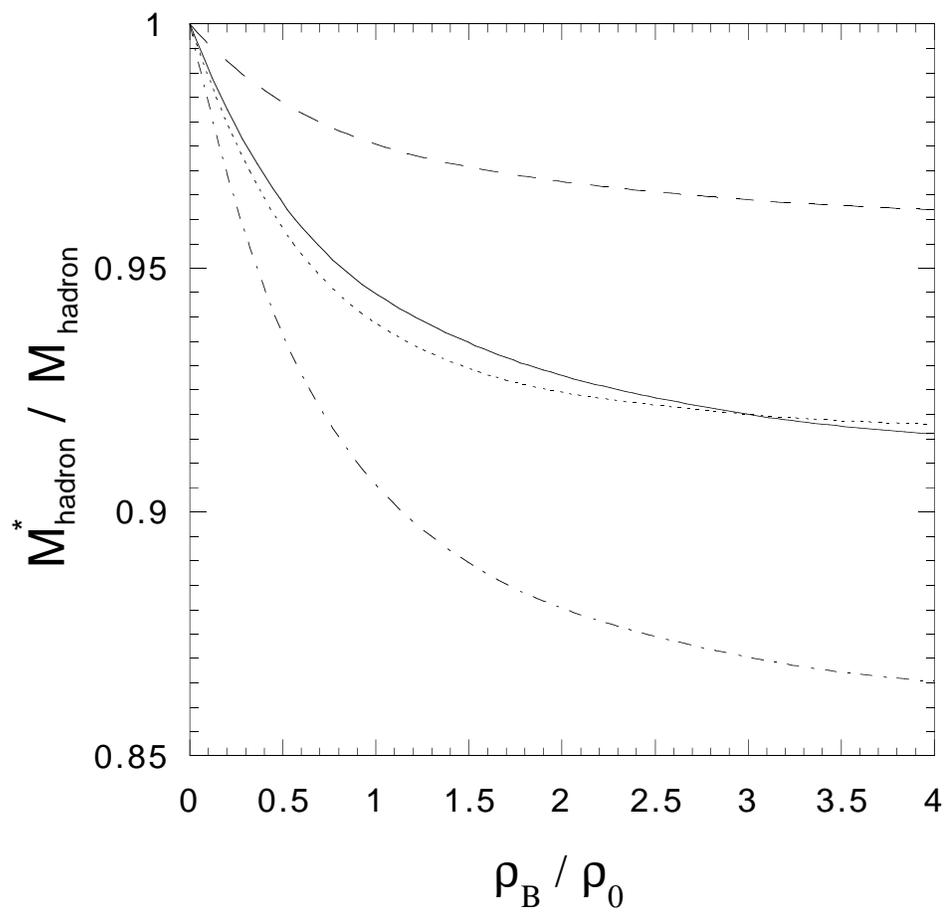
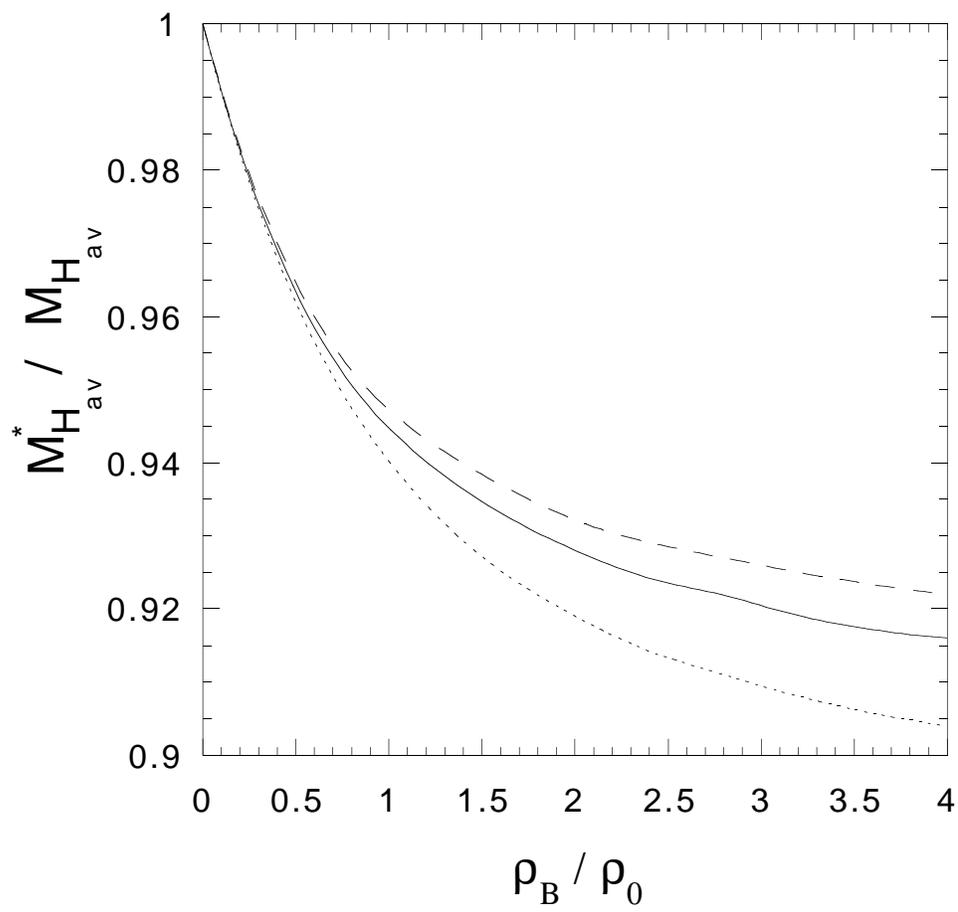
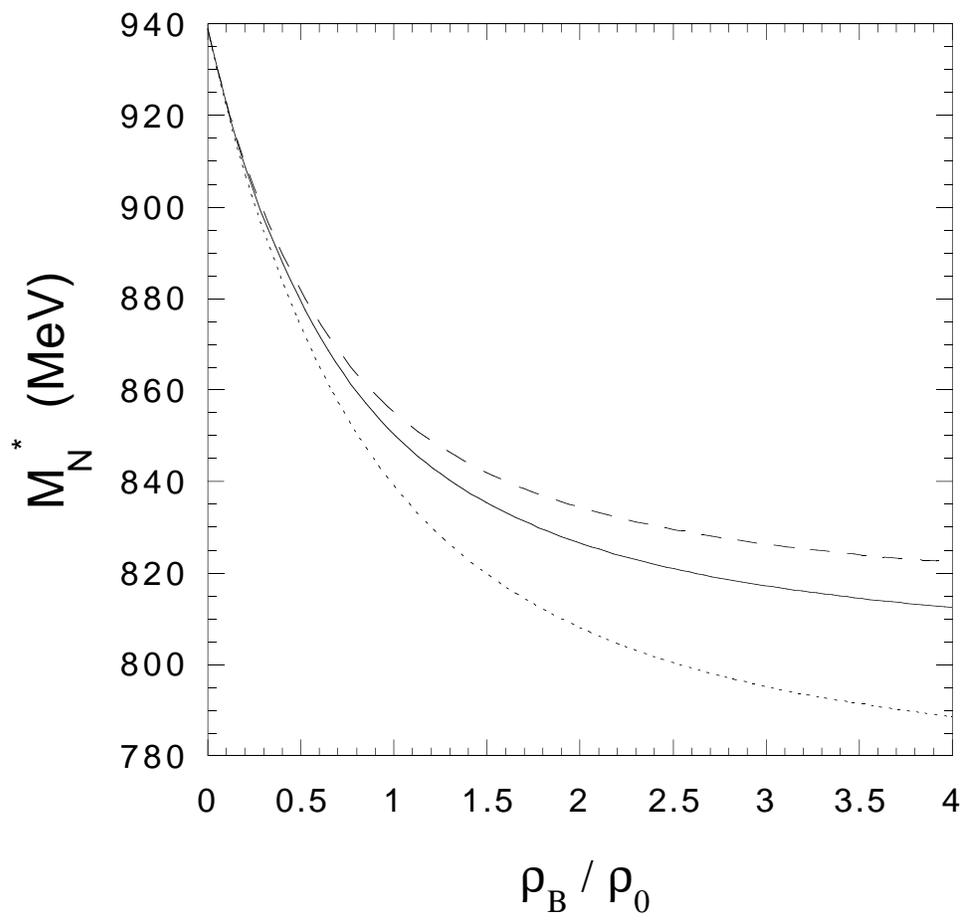
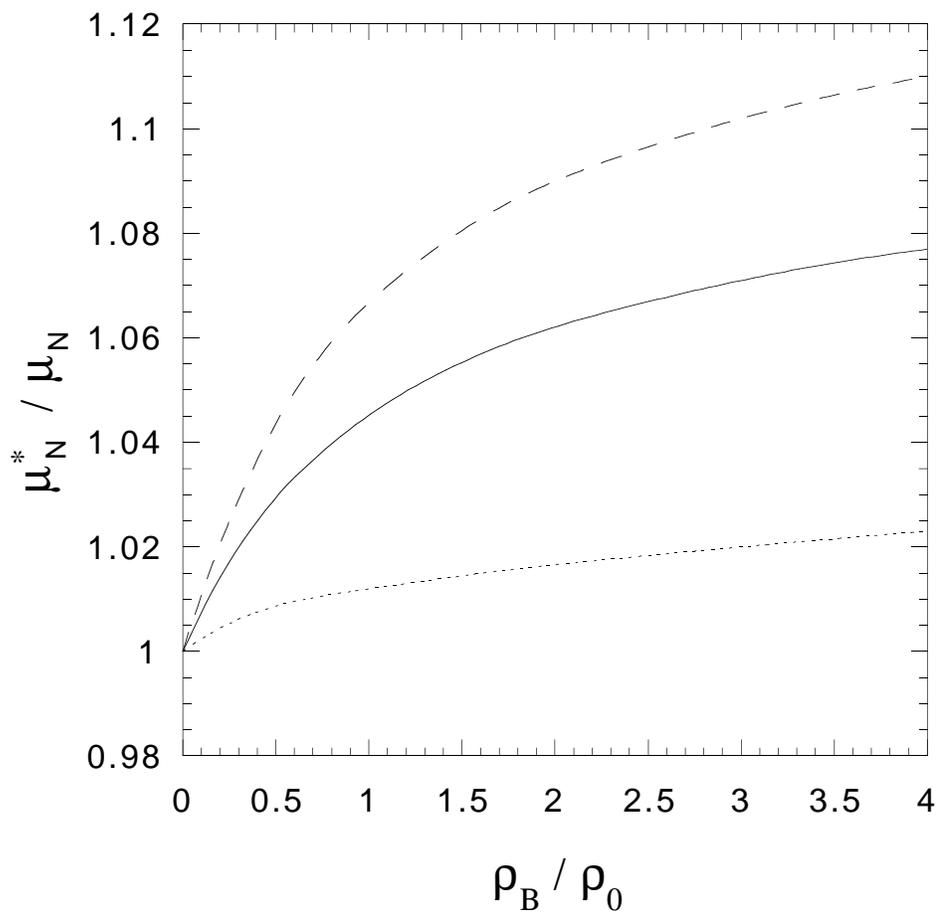
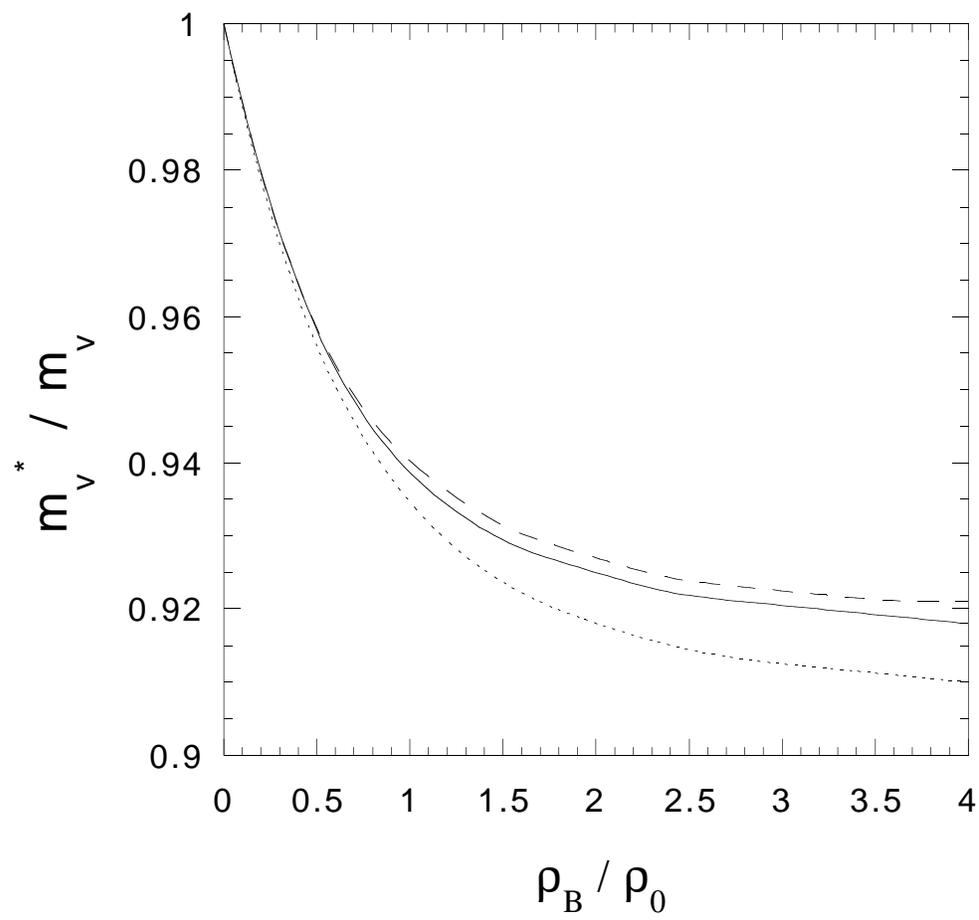
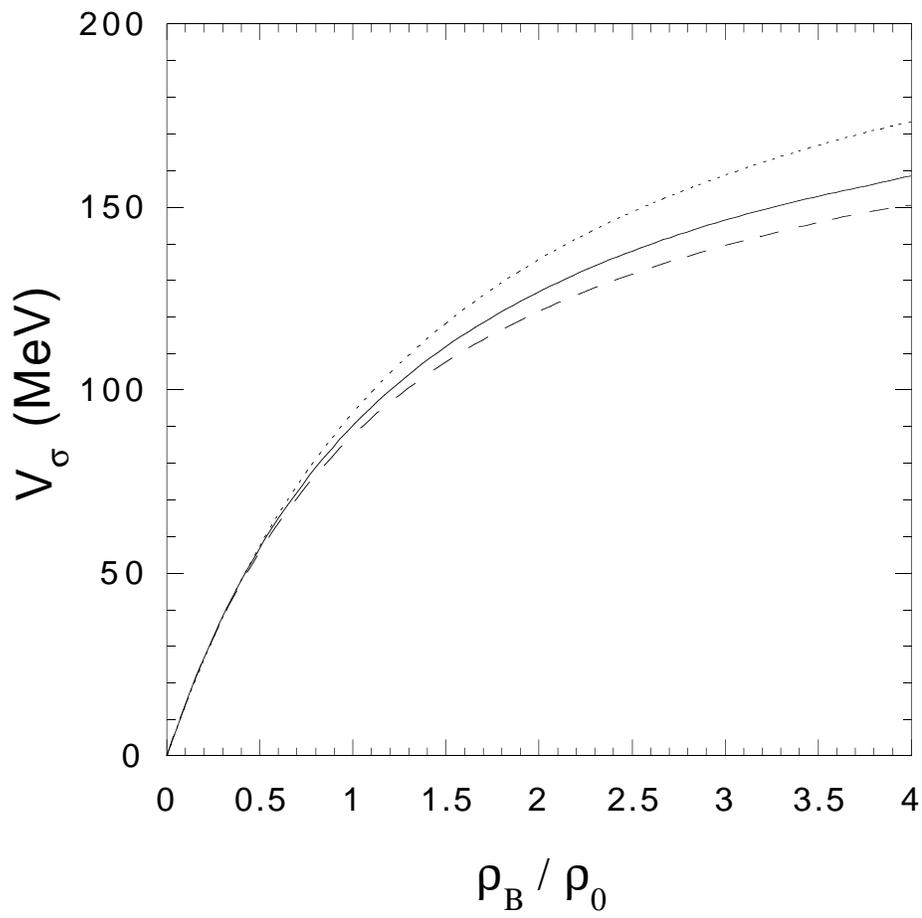
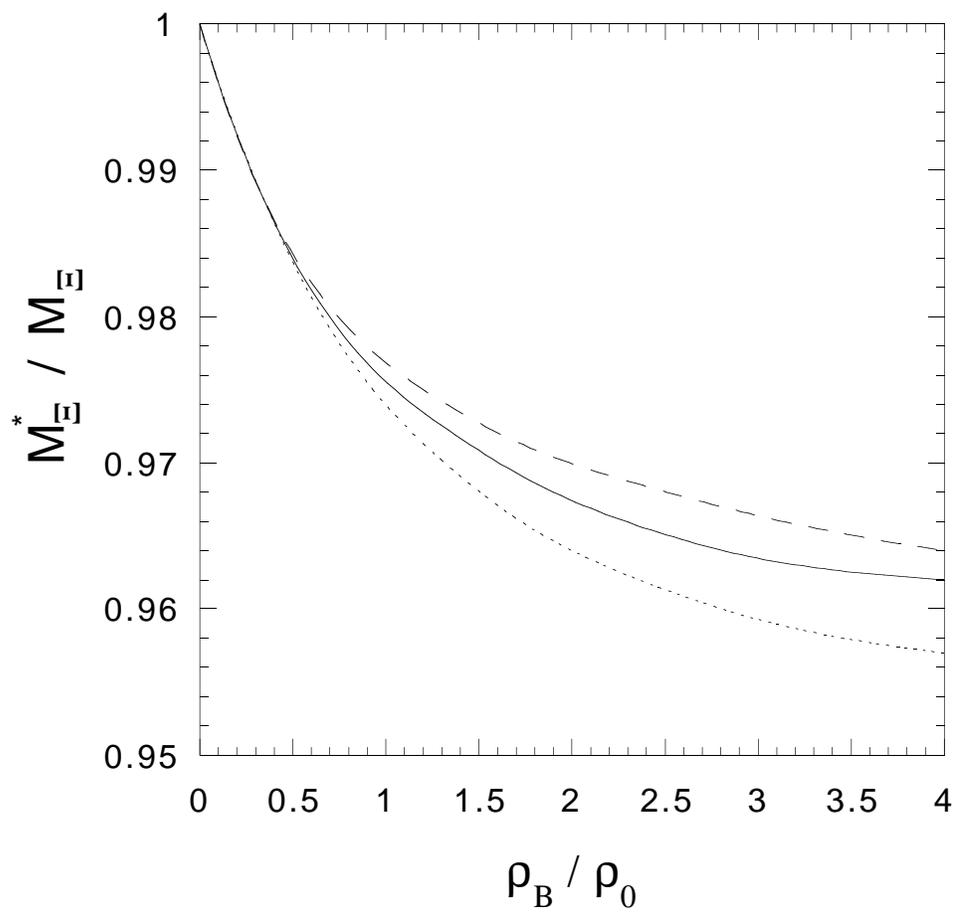
\end{document}